\documentclass[aps,prb,reprint,groupedaddress]{revtex4-2}
\usepackage[dvipdfmx]{graphicx}
\usepackage{amsmath}
\usepackage{amssymb}
\usepackage{mathtools}
\usepackage{mathcomp}
\usepackage{float}
\usepackage{bm}
\usepackage{color}
\usepackage{braket}
\usepackage{ulem}

\begin{document}


\title{
Conversion of chiral phonons into magnons in ferromagnets and antiferromagnets
}


\author{Dapeng Yao}
\email[]{yao.dapeng@stat.phys.titech.ac.jp}
\affiliation{Department of Physics, Tokyo Institute of Technology, 2-12-1 Ookayama, Meguro-ku, Tokyo 152-8551, Japan}
\author{Shuichi Murakami}
\email[]{murakami@stat.phys.titech.ac.jp}
\affiliation{Department of Physics, Tokyo Institute of Technology, 2-12-1 Ookayama, Meguro-ku, Tokyo 152-8551, Japan}

\date{\today}

\begin{abstract}
Chiral phonons with atomic rotations converted into electron spins result in a change of spin magnetizations in crystals.
In this paper, we investigate a new conversion of chiral phonons into magnons both in ferromagnets and antiferromagnets by spin models with exchange and Dzyaloshinskii-Moriya interactions.
The atomic rotations in chiral phonons are treated as slow phonons, which modulate spin-spin interaction and induce time-dependent magnon excitations due to geometric effect within adiabatic approximation. We demonstrate that a non-trivial change of the number of magnons requires breaking of the spin-rotation symmetry around the spin quantization axis.
As a result, the clockwise and counterclockwise chiral phonons induce a change of the magnon number with opposite signs, which corresponds to an increasing or decreasing spin magnetizaton due to the chiral nature of the atomic rotations.
In particular, in antiferromagnets, the modulation of magnons due to chiral phonons generates a non-zero net magnetization by the proposed effect, which is expect to be observed in experiments.
\end{abstract}

\maketitle

\section{introduction}
Coupling between electron spins and rotational motions enables various conversion phenomena, such as the Einstein-de Haas~\cite{EdH} and the Barnett effect~\cite{Barnett}, in which the mechanical rotation of the crystal acts on the electron spin as an effective magnetic field~\cite{SpRo1,SpRo2,SpRo3}.
Recently, chiral phonons, in which atomic motions have rotational modes, have been widely studied both in their theoretical formulations~\cite{PAM,CPthe,CPnonsym,kgmCP,PMAwoTRS,CPC4,FranssonCP} and in experimental observations~\cite{CHexp,CHobs_alpha}. 
On the other hand, these rotational motions of atoms around their equilibrium positions hold a chiral nature. 

In previous works, conversions of rotational motions of chiral phonons into electrons are theoretically studied by an adiabatic method. Because the masses of atoms are much larger than those of electrons, the rotations of atoms are slow, and the motions of electrons adiabatically respond to the atomic rotations. In this case, the contribution from chiral phonons can be treated by the Berry phase method~\cite{BerryPhase,geomterm}.
Under the setup of slow chiral phonons, one can show that chiral phonons generate a spin magnetization via spin-orbit coupling~\cite{CPmagHamada}. Moreover, in a chiral system with a helical crystal structure, in-plane chiral phonons induce a current along the helical axis, and the current induced by slow chiral phonons can have topological nature~\cite{CPcurrent}. These previous studies shed light on the conversion of chiral in paramagnetic electron systems and show that chiral phonons behaves as an effective magnetic field.

On the other hand, spin waves, which are also dubbed as magnons, are collective propagations of precessional motions of magnetic moments~\cite{BdG1,BdG2}, and they govern many unconventional transport phenomena including thermal Hall and spin Nernst effects~\cite{HEmag1,HEmag2,HEmag3,HEmag4,HEmag5,MHEexp,MHEthe} in magnetic systems. A spin-wave model on a pyrochlore lattice with Dzyaloshinskii-Moriya (DM) interaction has been proposed to investigate the magnon Hall effect in an insulating collinear ferromagnet Lu$_2$V$_2$O$_7$~\cite{MHEexp}. 
Similar physics is seen in a spin-wave model on a two-dimensional (2D) kagome lattice~\cite{MHEthe}, which corresponds to a (111) slice of the three-dimensional pyrochlore lattice. Accordingly, with the further understandings of chiral phonons, one can predict that the 2D kagome lattice can carry chiral phonons once the inversion symmetry is broken, in which clockwise (CW) and counterclockwise (CCW) rotational modes can appear at valley points connected by time reversal~\cite{kgmCP}. 
In addition, a circular polarized phonon mode can also appear on a 2D honeycomb lattice, which has been already observed by experiments~\cite{CHexp}.
Therefore, one can naturally expect conversions between chiral phonons and magnons in ferromagnets and antiferromagnets. 

In this paper, we find a new phenomenon of a conversion of chiral phonons into magnons both in ferromagnets and antiferromagnets with exchange and DM interactions. The atomic rotations in chiral phonons are treated as slow phonons, which modulate spin-spin interactions and induce time-dependent magnon excitations due to geometric effect within the adiabatic approximation.
As a result, a non-zero time average of a change of the magnon number due to the geometric effect appear and result in a change of spin magnetization both in ferromagnets and antiferromagnets if the spin-rotation symmetry is broken.
Intriguingly, the chiral phonons with CW and CCW modes induce a change in the number of magnons with opposite signs. Namely, they either increase or decrease the number of magnons due to the chiral nature of the atomic rotations.
This new phenomenon predicted in this paper is a magnetic analogue of the previous studies in an itinerant paramagnet~\cite{CPmagHamada,CPcurrent}. It suggests that in addition to electron spins in a paramagnet, magnons in a ferromagnet or an antiferromagnet also have a geometric effect generated by chiral phonons. In the case of magnons, we found that spin anisotropy plays an essential role, and the proposed effect is universal in a wide range of magnets. Compared with ferromagnetic magnons, antiferromagnetic magnons change their numbers such that the net spin magnetization no longer vanishes since chiral phonons work as an effective magnetic field. It provides the possibility to detect the proposed effect in antiferromagnets.

The remainder of the paper is organized as follows. In Sec.~\ref{secII}, we study the conversion of chiral phonons into magnons on the 2D ferromagnetic kagome lattice. Section~\ref{secIII} presents the conversion of chiral phonons into magnons on the 2D antiferromagnetic honeycomb lattice. Finally, we conclude our results in Sec.~\ref{secIV}. The details of our calculations on magnonic systems are placed in Appendix.

\section{conversion of chiral phonons into magnons in ferromagnets}
\label{secII}
In this section, we first study the conversion of chiral phonons into magnons in ferromagnets. We introduce a spin model with exchange and DM interactions, and also consider the chiral phonons, which modulate the spin-spin interactions and induce time-dependent magnon excitations.

\subsection{Ferromagnetic spin model on 2D kagome lattice}
\begin{figure}[htb]
\begin{center}
\includegraphics[clip,width=8.5cm]{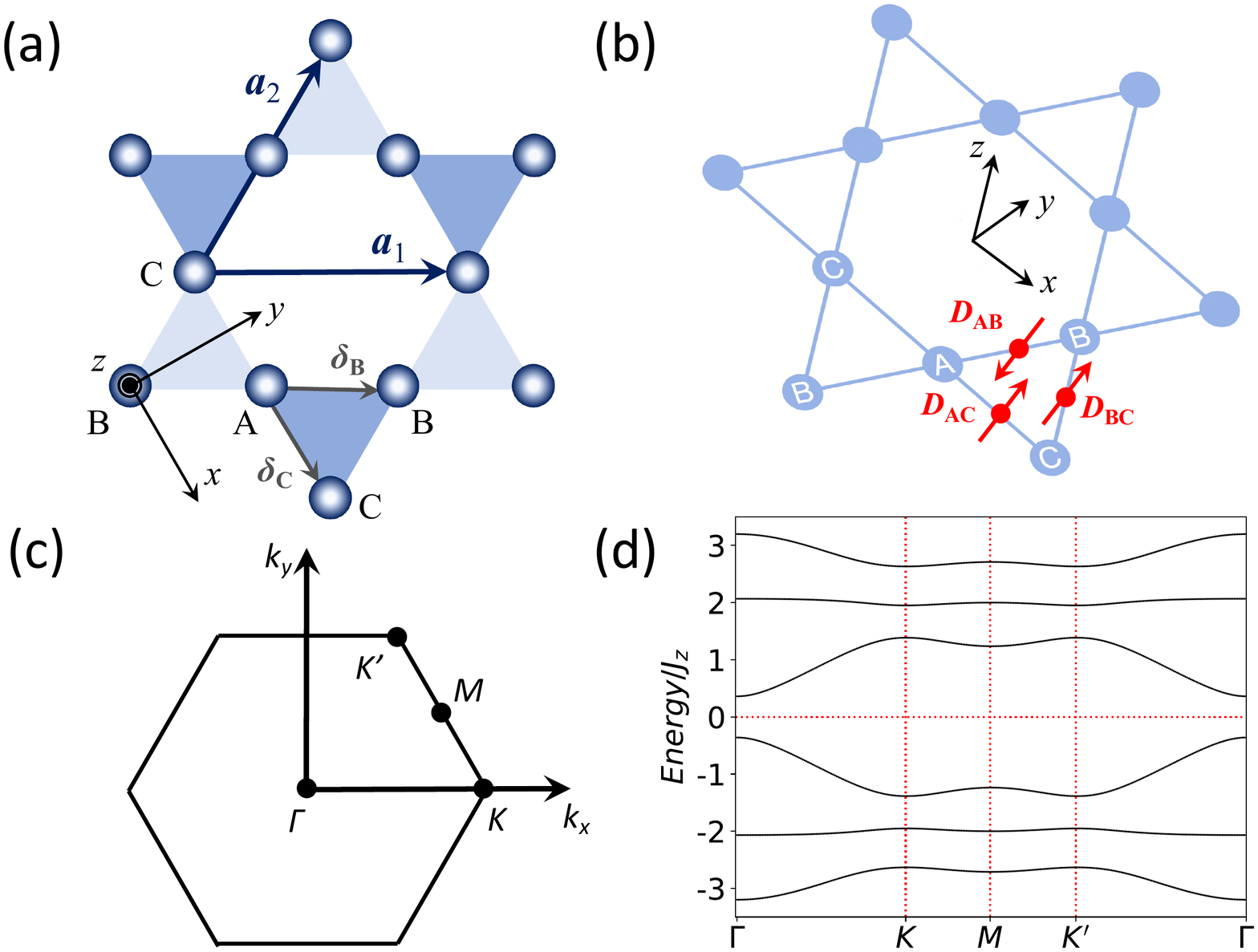}
\end{center}
\caption{
(Color online). Two-dimensional kagome lattice. (a) Three sublattices A, B, and C are placed at each corner of the triangles with lattice vectors $\bm{a}_1$ and $\bm{a}_2$. In each unit cell, the vectors $\bm{\delta}_B$ and $\bm{\delta}_C$ label the relative positions of the sublattice B and C to the sublattice A. (b) Dzyaloshinskii-Moriya vectors between two nearest-neighboring atoms are perpendicular to the $xy$ plane and represented by red (blue) arrows at the midpoints of two sublattices. To break the inversion symmetry, DM vectors and exchange parameters are set to be different between the adjacent sites. The DM vectors and exchange parameters are marked as $\bm{D}_{ij}$, $J_x$, $J_y$, $J_z$ in the red triangles and $\bm{D}_{ij}'$, $J_x'$, $J_y'$, $J_z'$ in the blue triangles.
(c) The first Brillouin zone with the high-symmetry points for the kagome lattice. (d) Magnon band structure from the spin-wave model with $J_z=J_z'=1$, $J_x=0.8$, $J_y=0.5$, $J_x'=0.85$, $J_y'=0.52$, $\lambda=0.4$, $S=1/2$, and $\tilde{H}=0$. There are three positive-definite bands above $E=0$, and three bands below $E=0$ are their copies.}
\label{kagome}
\end{figure}

We start from a quantum spin model with DM interaction on the 2D kagome lattice consisting of three sublattices A, B and C as shown in Fig.~\ref{kagome}(a). Here, the $z$ axis is set to be perpendicular to the 2D kagome-lattice plane, and $\bm{a}_1=a(1/2,\sqrt{3}/2)$ and $\bm{a}_2=a(-1/2,\sqrt{3}/2)$ are the primitive vectors of the lattice with the lattice constant $a$, and the vectors $\bm{\delta}_A=(0,0)$, $\bm{\delta}_B=a(1/4,\sqrt{3}/4)$ and $\bm{\delta}_C=a(1/2,0)$ label the relative positions of the sublattices A, B, and C with respect to the sublattice A. We notice that the bonds connecting adjacent sites form two types of triangles which are marked with red and blue in Fig.~\ref{kagome}(b).
The ferromagnetic spin Hamiltonian is given by
\begin{align}
\label{H0}
H_0=H_{\rm{ex}}+H_{\rm{DM}}+H_{\rm{mag}},
\end{align} 
where the first term
\begin{align} 
H_{\rm{ex}}=-\sum_{\braket{i,j}}\left(J_{ij}^xS_i^xS_j^x+J_{ij}^yS_i^yS_j^y+J_{ij}^zS_i^zS_j^z\right)
\end{align}
is the anisotropic XYZ spin model with different exchange parameters $J_{ij}^x$, $J_{ij}^y$ and $J_{ij}^z$ between electron spins located at the adjacent sites $i$ and $j$.
Here we suppose that the exchange parameters are different between the two types of the triangular unit cells. The exchange parameters are denoted by $J_{ij}^x=J_x$, $J_{ij}^y=J_y$, and $J_{ij}^z=J_z$ for the red triangle and $J_{ij}^x=J_x'$, $J_{ij}^y=J_y'$, and $J_{ij}^z=J_z'$ for the blue triangle. We set them to be different in order to break inversion symmetry, which is needed to have chiral phonons in the 2D kagome-lattice model~\cite{kgmCP}. If we set the exchange parameter to be isotropic $J_{ij}^x=J_{ij}^y=J_{ij}^z$, the first term $H_{\rm{ex}}$ reduces to an isotropic Heisenberg model.

The second term of the Hamiltonian in Eq.~(\ref{H0}) is the DM interaction which is given by
\begin{align}
\label{DMterm}
H_{\rm{DM}}=\sum_{\braket{i,j}}\bm{D}_{ij}\cdot\left(\bm{S}_i\times\bm{S}_j\right).
\end{align} 
Here, $\bm{D}_{ij}$ represents the DM interaction between the nearest-neighbor sites $i$ and $j$, and $\bm{D}_{ij}=-\bm{D}_{ji}$. This term is allowed since the midpoint between them is not a center of inversion. Since the directions of DM vectors are strongly constrained by the crystal symmetry, with the help of Moriya's rules~\cite{Moriya,MHEexp,MHEthe}, the DM vectors between each pair of nearest-neighboring sites can be obtained as
\begin{align}
\bm{D}_{AC}=\bm{D}_{CB}=\bm{D}_{BA}=\frac{2D}{\sqrt{6}}\hat{\bm{e}_z}, 
\end{align}
for the red triangles and
\begin{align}
\bm{D}'_{AC}=\bm{D}'_{CB}=\bm{D}'_{BA}=\frac{2D'}{\sqrt{6}}\hat{\bm{e}_z}, 
\end{align}
for the blue triangles. Here, these DM vectors have only the $z$ component as shown in Fig.~\ref{kagome}(b) with $D$ and $D'$ being the DM constants.

The third term of the Hamiltonian in Eq.~(\ref{H0})
\begin{align}
\label{Hmag}
H_{\rm{mag}}=-g\mu_B\tilde{H}\sum_iS^z_i
\end{align}
represents the coupling with an external magntic field $\tilde{\bm{H}}=\tilde{H}\hat{\bm{e}}_z$ along the $z$ axis, where $g$ and $\mu_B$ denote the $g$ factor of electrons and the Bohr's magneton, respectively. It is well known that in this spin model, the ground state is a collinear ferromagnet, and it is stable against the DM interaction. Here, since the component of the DM vector perpendicular to the direction of the external magnetic field does not contribute to the spin-wave Hamiltonian~\cite{MHEexp}, only the $z$ component of the DM vector $D_{ij}^z$ should be retained in this model. Therefore, the second term Eq.~(\ref{DMterm}) for the DM interaction can be expressed as  $H_{\rm{DM}}=\sum_{\braket{i,j}}D_{ij}^z\left(S_i^xS_j^y-S_i^yS_j^x\right)$.

Under the external magnetic field along the $z$ axis, we take the $z$ axis as a quantization axis of electron spins.
To obtain the magnon spectrum, we introduce the ladder operators $S_i^{\pm}=S_i^x\pm iS_i^y$.
Therefore, the anisotropic spin Hamiltonian is written as
\begin{align}
H_0=&-\frac{1}{4}\sum_{\braket{i,j}}\Bigg\{4J_{ij}^zS_i^zS_j^z+(J_{ij}^x-J_{ij}^y)(S_i^+S_j^++S_i^-S_j^-) \nonumber \\
        &+(J_{ij}^x+J_{ij}^y-i2D_{ij}^z)S_i^+S_j^-+(J_{ij}^x+J_{ij}^y+i2D_{ij}^z)S_i^-S_j^+\Bigg\} \nonumber \\
        &-g\mu_B\tilde{H}\sum_{i}S_i^z.
\end{align}
Then we apply the Holstein-Primakoff transformation
\begin{align}
\label{HP}
&S_i^z=S-a_i^{\dagger}a_i,  \nonumber \\
&S_i^+=\left(2S-a_i^{\dagger}a_i\right)^{1/2}a_i\approx\sqrt{2S}\left(a_i-\frac{1}{4S}a_i^{\dagger}a_ia_i\right), \nonumber \\ 
&S_i^-=a_i^{\dagger}\left(2S-a_i^{\dagger}a_i\right)^{1/2}\approx\sqrt{2S}\left(a_i^{\dagger}-\frac{1}{4S}a_i^{\dagger}a_i^{\dagger}a_i\right),
\end{align}
where $a_i^{\dagger}(a_i)$ is the creation (annihilation) operator of magnons, and $S$ represents the electron spin. When $S$ is large, the approximations in Eq.~(\ref{HP}) become good. By using this transformation and neglecting the second terms in Eq.~(\ref{HP}), the spin Hamiltonian is finally expressed as
\begin{align}
\label{bosonH0}
H_0=&-\frac{S}{2}\sum_{\braket{i,j}}(J_{ij}^x-J_{ij}^y)\left(a_i^{\dagger}a_j^{\dagger}+a_ia_j\right) \nonumber \\
       &-\frac{S}{2}\sum_{\braket{i,j}}J_{ij}^D\left(e^{i\phi_{ij}}a_i^{\dagger}a_j+e^{-i\phi_{ij}}a_j^{\dagger}a_i\right) \nonumber \\
       &+\sum_{i}(4J_{ij}^zS+g\mu_B\tilde{H})a_i^{\dagger}a_i,
\end{align}
where we introduce $J_{ij}^D=\sqrt{(J_{ij}^x+J_{ij}^y)^2+4(D_{ij}^z)^2}$ and $\phi_{ij}=\tan^{-1}\bigg(2D_{ij}^z/(J_{ij}^x+J_{ij}^y)\bigg)$, and the number 4 represents the number of the nearest-neighbor atoms. Thus, a quadratic form for the bosonic Hamiltonian has been finally obtained from the anisotropic spin model.

Here, we set $J_z=J_z'$ for simplicity, while the parameters $J_x$, $J_y$, and $D$ for the red triangles and $J_x'$, $J_y'$, and $D'$ for the blue triangles, are different. We also set $\lambda=\frac{D}{J_x+J_y}=\frac{D'}{J_x'+J_y'}$ for simplicity to guarantee that the angle $\phi_{ij}$ is common between the two types of triangles.
By means of the Fourier transformation
\begin{align}
\label{Fourier}
&a_{\bm{R}+\bm{\delta}_m}=\frac{1}{\sqrt{\mathcal{N}}}\sum_{\bm{k}}e^{-i\bm{k}\cdot(\bm{R}+\bm{\delta}_m)}a_{m,\bm{k}}, \nonumber \\
&a_{\bm{R}+\bm{\delta}_m}^{\dagger}=\frac{1}{\sqrt{\mathcal{N}}}\sum_{\bm{k}}e^{i\bm{k}\cdot(\bm{R}+\bm{\delta}_m)}a_{m,\bm{k}}^{\dagger},
\end{align}
where $m=$A, B, and C denoting the three sublattices, $\mathcal{N}$ is the total particle numbers within the finite size of the crystal, $\bm{R}$ represents the coordinate of the site, $\bm{k}$ denotes the wave vector, and $a_{m,\bm{k}}^{\dagger}(a_{m,\bm{k}})$ is the creation (annihilation) operator of the Bloch state with $\bm{k}$ being the Bloch wavevector. Then, Eq.~(\ref{bosonH0}) becomes a bosonic Bogoliubov-de Gennes (BdG) Hamiltonian,
\begin{align}
\label{BdG}
H_0=\frac{1}{2}\sum_{\bm{k}}
\begin{pmatrix}
\bm{\beta}_{\bm{k}}^{\dagger}, & \bm{\beta}_{-\bm{k}}
\end{pmatrix}
\mathcal{H}_0({\bm{k}})
\begin{pmatrix}
\bm{\beta}_{\bm{k}} \\
\bm{\beta}_{-\bm{k}}^{\dagger}
\end{pmatrix},
\end{align}
where $\bm{\beta}_{\bm{k}}^{\dagger}\equiv\left(a_{A,\bm{k}}^{\dagger},a_{B,\bm{k}}^{\dagger},a_{C,\bm{k}}^{\dagger}\right)$ represents the boson creation operators including the three internal degrees of freedom. The Bloch Hamiltonian of the spin model in the momentum space reads
\begin{align}
\mathcal{H}_0(\bm{k})=
\begin{pmatrix}
\mathcal{A}_{0,\bm{k}} & \mathcal{A}_{1,\bm{k}} & \mathcal{A}_{2,\bm{k}} & 0 & \mathcal{B}_{1,\bm{k}} & \mathcal{B}_{2,\bm{k}} \\
\mathcal{A}^*_{1,\bm{k}} & \mathcal{A}_{0,\bm{k}} & \mathcal{A}_{3,\bm{k}} & \mathcal{B}^*_{1,\bm{k}} & 0 & \mathcal{B}_{3,\bm{k}} \\
\mathcal{A}^*_{2,\bm{k}} & \mathcal{A}^*_{3,\bm{k}} & \mathcal{A}_{0,\bm{k}} & \mathcal{B}^*_{2,\bm{k}} & \mathcal{B}^*_{3,\bm{k}} & 0 \\
0 & \mathcal{B}^*_{1,-\bm{k}} & \mathcal{B}^*_{2,-\bm{k}} & \mathcal{A}_{0,-\bm{k}} & \mathcal{A}^*_{1,-\bm{k}} & \mathcal{A}^*_{2,-\bm{k}} \\
\mathcal{B}_{1,-\bm{k}} & 0 & \mathcal{B}^*_{3,-\bm{k}} & \mathcal{A}_{1,-\bm{k}} & \mathcal{A}_{0,-\bm{k}} & \mathcal{A}^*_{3,-\bm{k}} \\
\mathcal{B}_{2,-\bm{k}} & \mathcal{B}_{3,-\bm{k}} & 0 & \mathcal{A}_{2,-\bm{k}} & \mathcal{A}_{3,-\bm{k}} & \mathcal{A}_{0,-\bm{k}} \\
\end{pmatrix},
\end{align}
where
\begin{align}
&\mathcal{A}_{0,\bm{k}}=4J_zS+g\mu_B\tilde{H}, \nonumber \\
&\mathcal{A}_{1,\bm{k}}=-\frac{S}{2}e^{-i\phi}\left\{J_De^{-iK_{AB}}+J'_De^{iK_{AB}}\right\}, \nonumber \\
&\mathcal{A}_{2,\bm{k}}=-\frac{S}{2}e^{i\phi}\left\{J_De^{-iK_{AC}}+J'_De^{iK_{AC}}\right\}, \nonumber \\
&\mathcal{A}_{3,\bm{k}}=-\frac{S}{2}e^{-i\phi}\left\{J_De^{-iK_{BC}}+J'_De^{iK_{BC}}\right\}, \nonumber \\
&\mathcal{B}_{1,\bm{k}}=-S\left\{(J_x-J_y)e^{-iK_{AB}}+(J'_x-J'_y)e^{iK_{AB}}\right\}, \nonumber \\
&\mathcal{B}_{2,\bm{k}}=-S\left\{(J_x-J_y)e^{-iK_{AC}}+(J'_x-J'_y)e^{iK_{AC}}\right\}, \nonumber \\
&\mathcal{B}_{3,\bm{k}}=-S\left\{(J_x-J_y)e^{-iK_{BC}}+(J'_x-J'_y)e^{iK_{BC}}\right\},
\end{align}
with 
$J_D=\sqrt{(J_x+J_y)^2+\frac{8D^2}{3}}$,  
$\phi=\tan^{-1}\left(\frac{4D}{\sqrt{6}(J_x+J_y)}\right)$, 
$K_{AB}=(k_x+\sqrt{3}k_y)a/4$, $K_{AC}=k_xa/2$, and $K_{BC}=(k_x-\sqrt{3}k_y)a/4$.
Here $\mathcal{A}_{0,\bm{k}}$ denotes the on-site energy of magnons for which sublattices A, B, and C have no difference, and $\mathcal{A}_{1,\bm{k}}$, $\mathcal{A}_{2,\bm{k}}$, and $\mathcal{A}_{3,\bm{k}}$ ($\mathcal{B}_{1,\bm{k}}$, $\mathcal{B}_{2,\bm{k}}$, and $\mathcal{B}_{3,\bm{k}}$) represent the nearest neighbor interaction terms between the A-B, A-C, and B-C sublattices, respectively. We notice that both the exchange and the DM interaction contribute to $\mathcal{A}_{i,\bm{k}}$, and that $\mathcal{B}_{i,\bm{k}} (i=1,2,3)$ has only the contributions from the $x$ and $y$ components of the exchange interaction.
The details of the diagonalization for the bosonic BdG Hamiltonian are given in Appendix~\ref{A}. The magnon bands can be obtained as shown in Fig.~\ref{kagome}(d) with parameters $J_z=J_z'=1$, $J_x=0.8$, $J_y=0.5$, $J_x'=0.85$, $J_y'=0.52$, $\lambda=0.4$, and $S=1/2$ without the external magnetic field ($\tilde{H}=0$). Here the magnon dispersions are plotted along the high-symmetry points in Fig.~\ref{kagome}(c). The bands above $E=0$ are the magnon bands which are physically meaningful, and the bands below $E=0$ are copies of the magnon bands because of the BdG form of the Hamiltonian, as discussed in Appendix~\ref{A}. 
The magnon spectrum is gapped, meaning that the ground state of this system is still ferromagnetic under the anisotropic exchange and the DM interaction. We notice that the bands at the $\bm{K}$ and $\bm{K}'$ point are not identical since we set that $J_x$, $J_y$, and $D$ are different from $J_x'$, $J_y'$, and $D'$ here, by which the inversion symmetry are broken.

\subsection{Modulation of spin-spin interactions by chiral phonons on 2D kagome lattice}
To find a model which can well describe ferromagnetic systems with chiral phonons, we add a perturbation term corresponding to chiral phonons into the static spin Hamiltonian constructed in the previous subsection. 

\begin{figure}[htb]
\begin{center}
\includegraphics[clip,width=9cm]{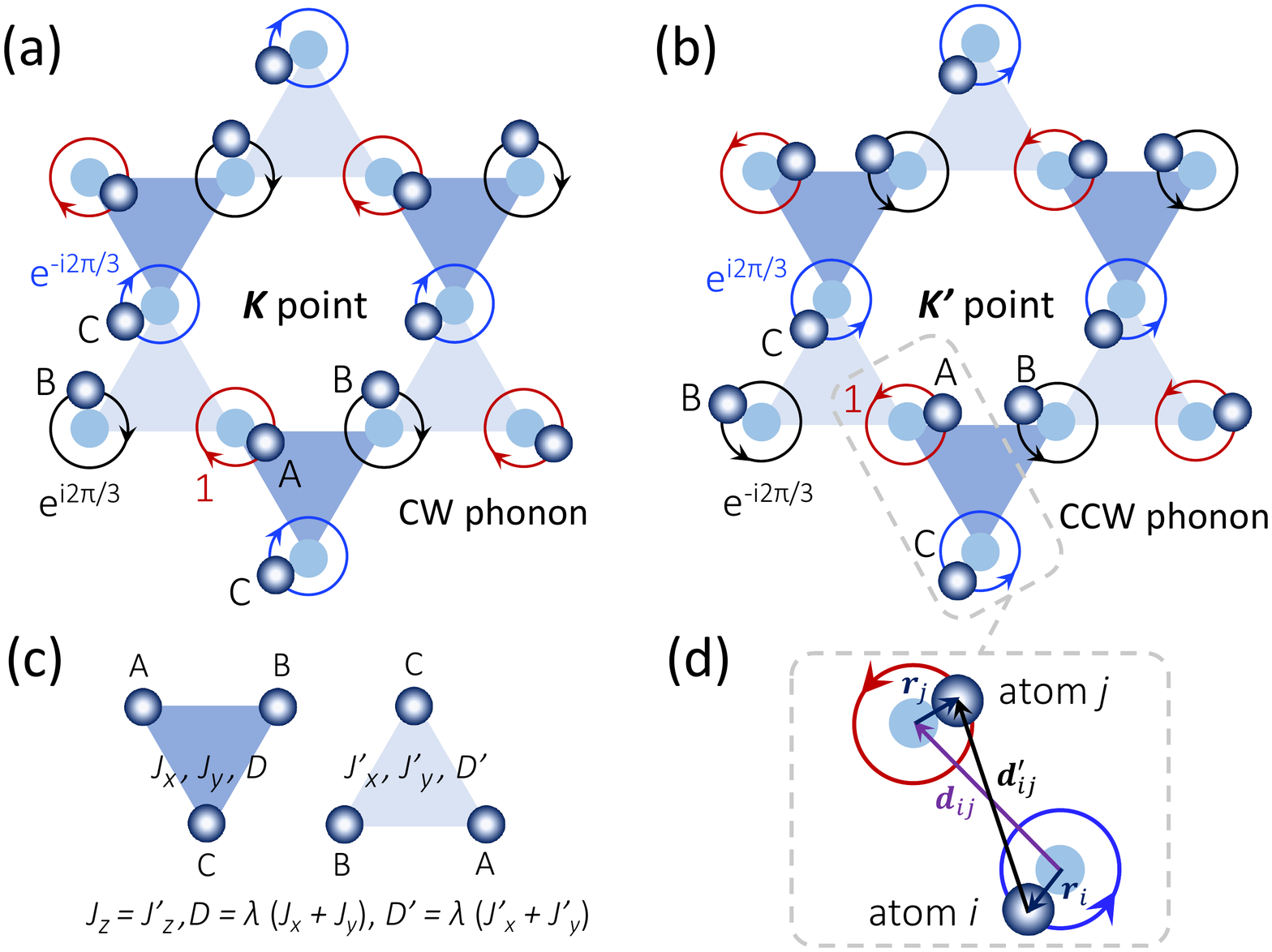}
\end{center}
\caption{
(Color online). Two kinds of chiral phonon modes. Three atoms in each unit cell rotate in the $xy$ plane with a certain phase correlation. The blue atoms rotate around their equilibrium positions with $\pm2\pi/3$ phase difference. (a) Clockwise (CW) phonon mode at $\bm{K}$ point. (b) Counterclockwise (CCW) phonon mode at $\bm{K}'$ point. (c) Parameters of the upward triangles and the downward triangles are set to be different, which breaks inversion symmetry. (d) Two atoms $i$ and $j$ with their displacement vectors $\bm{r}_i$ and $\bm{r}_j$ from their equilibrium positions. Their relative positions are $\bm{d}_{ij}$ and $\bm{d}'_{ij}$ corresponding to the cases without and with rotational motion, respectively.
}
\label{phonon_modes}
\end{figure}

In general, the microscopic local rotation of atoms around their equilibrium positions carries a phase difference between adjacent atoms. In the present model, since the exchange parameters $J_{ij}^a(a=x,y,z)$ and the DM vector $D_{ij}^z$ are proportional to the overlap integrals between the atomic orbitals involved, they depend on the distance between the two adjacent magnetic atoms. 
In the presence of chiral phonons, the exchange parameters and the DM vectors should be modulated because the distance between two adjacent magnetic atoms changes periodically with time due to atomic rotations. 
Figure~\ref{phonon_modes}(d) shows the rotational motions of the two nearest-neighboring atoms $i$ and $j$, where the displacement vectors relative to their equilibrium positions are set to be $\bm{r}_i(t)$ and $\bm{r}_j(t)$, which depend on the time $t$. It is reasonable to set the modulations of the spin-spin interactions to be proportional to the length change between the atoms given by $\bm{r}_{ij}(t)\cdot\left(\bm{d}_{ij}/a_0\right)$, where $\bm{r}_{ij}(t)=\bm{r}_j(t)-\bm{r}_i(t)$ and $a_0=a/2$ for the kagome lattice. Thus, the exchange parameters and the DM vectors are modulated as $J_{ij}^{\alpha}\rightarrow J_{ij}^{\alpha}-\frac{J_{ij}^{\alpha}}{a_0}\bm{r}_{ij}(t)\cdot(\bm{d}_{ij}/a_0)~(\alpha=x,y,z)$, and $D_{ij}^z\rightarrow D_{ij}^z-\frac{D_{ij}^z}{a_0}\bm{r}_{ij}(t)\cdot(\bm{d}_{ij}/a_0)$, respectively. Therefore, the parameter $J_{ij}^D$ which couples the exchange parameter $J_{ij}^x$, $J_{ij}^y$, and the DM vector $D_{ij}^z$ in the Eq.~(\ref{bosonH0}) should also be modulated as $J_{ij}^D\rightarrow J_{ij}^D-\frac{J_{ij}^D}{a_0}\bm{r}_{ij}(t)\cdot(\bm{d}_{ij}/a_0)$. Finally, the modulated spin Hamiltonian with chiral phonons can be expressed as
\begin{align}
\label{dH}
\delta H(t)=&-\frac{S}{2}\sum_{\braket{i,j}}\frac{J_{ij}^x-J_{ij}^y}{a_0^2}\bm{r}_{ij}(t)\cdot\bm{d}_{ij}\left(a_i^{\dagger}a_j^{\dagger}+a_ia_j\right) \nonumber \\
             &-\frac{S}{2}\sum_{\braket{i,j}}\frac{J_{ij}^D}{a_0^2}\bm{r}_{ij}(t)\cdot\bm{d}_{ij}\left(e^{i\phi_{ij}}a_i^{\dagger}a_j+e^{-i\phi_{ij}}a_j^{\dagger}a_i\right) \nonumber \\
             &-\frac{S}{2}\sum_{\braket{i,j}}\frac{J_{ij}^z}{a_0^2}\bm{r}_{ij}(t)\cdot\bm{d}_{ij}\left(a^{\dagger}_ia_i+a^{\dagger}_ja_j\right),
\end{align}
in which the last term can be cancelled if we set the parameter $J_z=J_z'$ as same as the one in the previous spin model without chiral phonons. Therefore the modulated spin Hamiltonian $\delta H(t)$ has no diagonal terms with the chiral phonons here. Later we will use Eq.~(\ref{dH}) with a certain phonon mode for the 2D kagome lattice.

It has been predicted that on the 2D kagome lattice, the chiral phonons can emerge at the valley $\bm{K}$ and $\bm{K}'$ points if the inversion symmetry is broken by introducing the different spring constants between the two types of triangles (blue and red ones in Fig.~\ref{kagome}(b)), and their vibration trajectories for the chiral phonons can be circular or elliptical in a 2D kagome lattice~\cite{kgmCP}. Here, as shown in Fig.~\ref{phonon_modes}(a) and (b), we consider two circular-polarized chiral-phonon modes, which appear at the $\bm{K}$ point with a CW rotational mode and at the $\bm{K}'$ point with a CCW rotational mode, and they are justified by the previous prediction~\cite{kgmCP}.
In Fig.~\ref{phonon_modes}(a), the displacement vectors of the sublattices A, B, and C for CW phonons are written as 
$\bm{r}_A^{\rm{cw}}=r_0(\cos\omega t,-\sin\omega t)$, $\bm{r}_B^{\rm{cw}}=r_0(\cos(\omega t-2\pi/3),-\sin(\omega t-2\pi/3))$, $\bm{r}_C^{\rm{cw}}=r_0(\cos(\omega t+2\pi/3),-\sin(\omega t+2\pi/3))$, and Fig.~\ref{phonon_modes}(b) shows that the displacement vectors for CCW phonons are $\bm{r}_A^{\rm{ccw}}=r_0(\cos\omega t,\sin\omega t)$, $\bm{r}_B^{\rm{ccw}}=r_0(\cos(\omega t+2\pi/3),\sin(\omega t+2\pi/3))$, $\bm{r}_C^{\rm{ccw}}=r_0(\cos(\omega t-2\pi/3),\sin(\omega t-2\pi/3))$ with the phonon frequency being $\omega$.
In this case, $\bm{r}_{ij}(t)\cdot\bm{d}_{ij}$ for the adjacent sublattices A-B, A-C, and B-C can be expressed as $\bm{r}_{AB}(t)\cdot\bm{\delta}_B=-\frac{\sqrt{3}ar_0}{2}\sin(\pm\omega t)$, $\bm{r}_{AC}(t)\cdot\bm{\delta}_C=-\frac{\sqrt{3}ar_0}{2}\sin(\pm\omega t+\pi/3)$, and $\bm{r}_{BC}(t)\cdot(\bm{\delta}_C-\bm{\delta}_B)=-\frac{\sqrt{3}ar_0}{2}\sin(\pm\omega t-\pi/3)$, where $\pm$ represents the CW and CCW phonons, respectively.

By using the Fourier transformation Eq.~(\ref{Fourier}), a Bloch Hamiltonian for the perturbation Eq.~(\ref{dH}) can be written in a quadratic form of a bosonic operators as
\begin{align}
\label{dHBdG}
\delta H(t)=\frac{1}{2}\sum_{\bm{k}}
\begin{pmatrix}
\bm{\beta}_{\bm{k}}^{\dagger}, & \bm{\beta}_{-\bm{k}}
\end{pmatrix}
\delta\mathcal{H}({\bm{k}},t)
\begin{pmatrix}
\bm{\beta}_{\bm{k}} \\
\bm{\beta}_{-\bm{k}}^{\dagger}
\end{pmatrix},
\end{align}
with
\begin{widetext}
\begin{eqnarray}
\delta\mathcal{H}(\bm{k},t)=
\begin{pmatrix}
0 & \delta \mathcal{A}_{1,\bm{k}}(t) & \delta \mathcal{A}_{2,\bm{k}}(t) & 0 & \delta \mathcal{B}_{1,\bm{k}}(t) & \delta \mathcal{B}_{2,\bm{k}}(t) \\
\delta \mathcal{A}^*_{1,\bm{k}}(t) & 0 & \delta \mathcal{A}_{3,\bm{k}}(t) & \delta \mathcal{B}^*_{1,\bm{k}}(t) & 0 & \delta \mathcal{B}_{3,\bm{k}}(t) \\
\delta \mathcal{A}^*_{2,\bm{k}}(t) & \delta \mathcal{A}^*_{3,\bm{k}}(t) & 0 & \delta \mathcal{B}^*_{2,\bm{k}}(t) & \delta \mathcal{B}^*_{3,\bm{k}}(t) & 0 \\
0 & \delta b^*_{1,-\bm{k}}(t) & \delta \mathcal{B}^*_{2,-\bm{k}}(t) & 0 & \delta \mathcal{A}^*_{1,-\bm{k}}(t) & \delta \mathcal{A}^*_{2,-\bm{k}}(t) \\
\delta \mathcal{B}_{1,-\bm{k}}(t) & 0 & \delta \mathcal{B}^*_{3,-\bm{k}}(t) & \delta \mathcal{A}_{1,-\bm{k}}(t) & 0 & \delta \mathcal{A}^*_{3,-\bm{k}}(t) \\
\delta \mathcal{B}_{2,-\bm{k}}(t) & \delta \mathcal{B}_{3,-\bm{k}}(t) & 0 & \delta \mathcal{A}_{2,-\bm{k}}(t) & \delta \mathcal{A}_{3,-\bm{k}}(t) & 0 \\
\end{pmatrix}.
\end{eqnarray}
\end{widetext}
Here the elements of the Bloch Hamiltonian for the perturbation are 
\begin{align}
\delta \mathcal{A}_{1,\bm{k}}(t)=&-\frac{S}{2}e^{-i\phi}\left\{\delta J_{D}e^{-iK_{AB}}+\delta J_{D}'e^{iK_{AB}}\right\}\sin(\pm\omega t), \nonumber \\
\delta \mathcal{A}_{2,\bm{k}}(t)=&-\frac{S}{2}e^{i\phi}\left\{\delta J_{D}e^{-iK_{AC}}+\delta J_{D}'e^{iK_{AC}}\right\}\sin(\pm\omega t-\frac{\pi}{3}), \nonumber \\
\delta \mathcal{A}_{3,\bm{k}}(t)=&-\frac{S}{2}e^{-i\phi}\left\{\delta J_{D}e^{-iK_{BC}}+\delta J_{D}'e^{iK_{BC}}\right\}\sin(\pm\omega t+\frac{\pi}{3}), \nonumber \\
\delta \mathcal{B}_{1,\bm{k}}(t)=&-\sqrt{3}S\left\{C_be^{-iK_{AB}}+C'_be^{iK_{AB}}\right\}\sin(\pm\omega t), \nonumber \\
\delta \mathcal{B}_{2,\bm{k}}(t)=&-\sqrt{3}S\left\{C_be^{-iK_{AC}}+C'_be^{iK_{AC}}\right\}\sin(\pm\omega t-\frac{\pi}{3}), \nonumber \\
\delta \mathcal{B}_{3,\bm{k}}(t)=&-\sqrt{3}S\left\{C_be^{-iK_{BC}}+C'_be^{iK_{BC}}\right\}\sin(\pm\omega t+\frac{\pi}{3}), 
\end{align}
where $\delta J_{D}=r_0\sqrt{(J_x+J_y)^2+\frac{8D^2}{3}}/a_0$, $\delta J_{D}'=r_0\sqrt{(J'_x+J'_y)^2+\frac{8D'^2}{3}}/a_0$, $C_b=\delta J_x-\delta J_y$, $C'_b=\delta J'_x-\delta J'_y$ with $\delta J_{\alpha}=J_{\alpha}r_0/a_0~(\alpha=x,y)$. Here $r_0$ denotes the amplitude of the atomic motions, and we set $r_0=0.1a_0$ in the following calculation as an example, which means the displacement is around $10\%$ of the lattice constant~\cite{CPmagHamada,CPcurrent}.
The details of our calculations will be shown later.

\subsection{Adiabatic magnon dynamics and symmetry constrains}
In our assumption of slow chiral phonons, the atomic rotations adiabatically affect the spin configurations of electrons so that an additional magnon dynamics is induced by chiral phonons. Here, we investigate the change of the number of the magnons based on the modulated spin model by means of Berry phase treatment, which has already been applied to electron systems for calculating orbital magnetization, spin magnetization and current under adiabatic process~\cite{geomterm,CPmagHamada,CPcurrent}. It starts from a periodically time-dependent Hamiltonian $H_t$ which changes slowly enough compared to the electronic energy scale. Let $\ket{\psi_n(t)}$ be an instantaneous eigenstate of the band $n$ at time $t$. The expectation value of the operator $\hat{X}$ that we are interested in under the adiabatic process can be obtained as 
\begin{align}
\label{Xt}
X(t)=\sum_n^{\rm{occ}}\left(X_n^{\rm{inst}}(t)+X_n^{\rm{geom}}(t)\right).
\end{align}
Here, the first term
\begin{align}
\label{inst}
X_n^{\rm{inst}}(t)\equiv\hbar\bra{\psi_n(t)}\hat{X}(t)\ket{\psi_n(t)}
\end{align}
is called an instantaneous term and the second term
\begin{align}
\label{geom}
X_n^{\rm{geom}}(t)\equiv\sum_{m(\neq n)}\left\{\frac{\hat{X}_{nm}(t)A_{mn}(t)}{E_n(t)-E_m(t)}+\rm{c.c}\right\}
\end{align}
is dubbed as a geometric term with the matrix elements $\hat{X}_{nm}(t)=\bra{\psi_n(t)}\hat{X}(t)\ket{\psi_m(t)}$ of the operator $\hat{X}(t)$, and the Berry phase $A_{mn}(t)=\bra{\psi_m(t)}(-i\partial_t)\ket{\psi_n(t)}$.
The formulation of Eq.~(\ref{Xt}) is for fermionic systems and it is expressed as the sum $\sum_n^{\rm{occ}}$ over all the occupied states at the absolute zero temperature. In bosonic systems, one should replace Eq.~(\ref{Xt}) with a corresponding formula with the Bose distribution function.

Since chiral phonons modulate the spin-spin interactions, the Bloch state of the modulated spin Hamiltonian will change with time. Here, we focus on the change of magnon numbers, which corresponds to the change of magnetizations. Within the adiabatic approximation, only the geometric term can give a non-trivial result because the geometric term records a path-dependent dynamics in terms of the Berry connection during an adiabatic process. 
In particular, the geometric terms of the CW and CCW phonons have the opposite paths in the phase space, which causes opposite changes of the number of magnons and reflects the phonon chirality. On the other hand, the instantaneous term represents the snapshot at time $t$, and it has no difference between the CW and CCW phonons unlike the geometric term. 

Now, we consider the number operator of magnons $\hat{N}$ as the operator $\hat{X}$. Then, we need to rewrite Eq.~(\ref{Xt}) to incorporate the Bose distribution function. To avoid the singularity of the Bose distribution function at zero temperature, we set the temperature to be nonzero and then Eq.~(\ref{geom}) should be revised accordingly by incorporating the Bose distribution function as
\begin{align}
\label{geomN}
&N_n^{\rm{geom}}(\tau) \nonumber \\
&=\frac{\hbar\omega}{S_{xy}}\sum_{m(\neq n)}\sum_{\bm k}\left\{\frac{\hat{N}_{nm}(\bm{k},\tau)A_{mn}(\bm{k},\tau)}{E_{n,\bm{k}}(\tau)-E_{m,\bm{k}}(\tau)}+\rm{c.c}\right\}f_{n,\bm{k}},
\end{align}
where 
\begin{align}
\label{Nnm}
\hat{N}_{nm}(\bm{k},\tau)=\bra{\psi_{n,\bm{k}}(\tau)}\hat{N}(\tau)\ket{\psi_{m,\bm{k}}(\tau)}
\end{align}
is the off-diagonal matrix element of the number operator of magnons, and 
\begin{align}
\label{Amn}
A_{mn}(\bm{k},\tau)=&\bra{\psi_{m,\bm{k}}(\tau)}(-i\partial_\tau)\ket{\psi_{n,\bm{k}}(\tau)} \nonumber \\
                  =&-i\frac{\bra{\psi_{m,\bm{k}}(\tau)}\partial_{\tau}\mathcal{H}(\bm{k},\tau)\ket{\psi_{n,\bm{k}}(\tau)}}{E_{n,\bm{k}}(\tau)-E_{m,\bm{k}}(\tau)}
\end{align}
is the matrix element of the Berry connection. Here, we introduce a dimensionless quantity $\tau=\omega t$ to replace the time $t$, $S_{xy}$ represents the size of the 2D unit cell within the $xy$ plane, and $f_{n,\bm{k}}=f(E_{n,\bm{k}})$ is the Bose distribution function for the $n$-th band. $N_n^{\rm{geom}}(\tau)$ as a function of $\tau$ becomes proportional to the phonon frequency $\omega$ because the Berry connection contains the time derivative terms.

The details of the calculation of the geometric term within the bosonic BdG formalism is discussed in Appendix~\ref{B}. We notice that the geometric term involves the off-diagonal elements of the number operator of magnons. If we consider an isotropic spin model, however, the number operator of magnons can be simultaneously diagonalized with the isotropic spin Hamiltonian because of the SU(2) spin-rotation symmetry. Therefore, the isotropy results in a zero geometric term of the number of magnons. That is why we introduce an anisotropic spin model in this paper in order to obtain a nonzero change of the magnon excitations induced by chiral phonons.

\subsection{Change of magnon numbers due to chiral phonons in ferromagnets}
\begin{figure*}[htb]
\begin{center}
\includegraphics[clip,width=18.5cm]{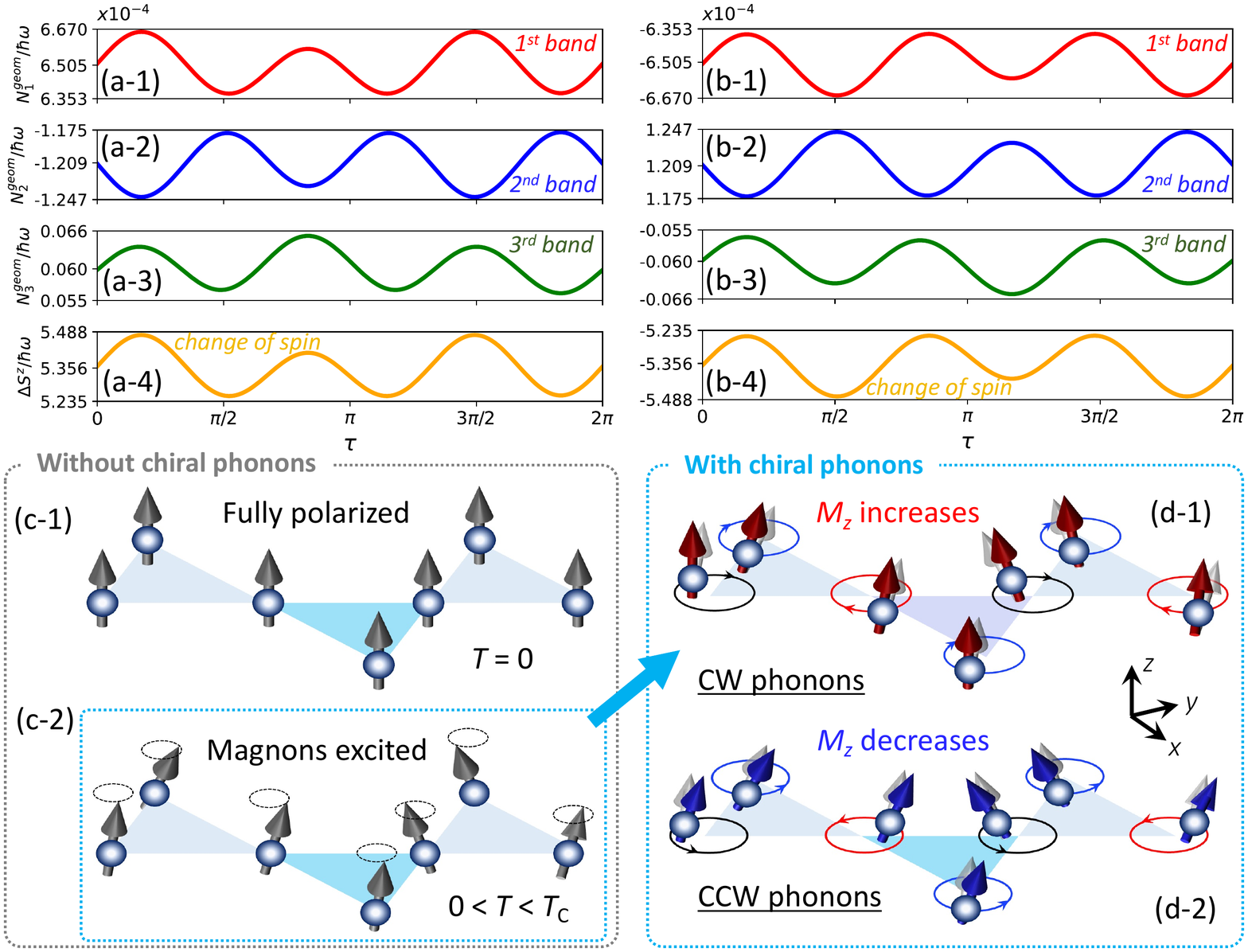}
\end{center}
\caption{
(Color online). The number of magnons varying with with time $t$ for one period on 2D kagome lattice. Here, $\tau=\omega t$ is the dimensionless time. (a)(b) Magnon excitation by (a) CW phonons and (b) CCW phonons. Here red, blue, and green lines represent the number of magnons from the first, second, and third bands, and orange lines represent their sum, which corresponds the change of spin $\Delta S^z$ due to chiral phonons. The middle number on the vertical axis represents the time average over one period $0\leq\tau\leq2\pi$. The parameters are set to be $J_z=J_z'=1$, $J_x=0.8$, $J_y=0.5$, $J_x'=0.85$, $J_y'=0.52$, $r_0=0.1a_0$, $\lambda=0.4$, $S=1/2$, and $\tilde{H}=0$. (c-1) and (c-2) show schematic figures of spin configuration without chiral phonons at  $T=0$ (fully polarized) and at $0<T<T_c$ (magnon excited), respectively. (d-1) and (d-2) show the changed spin configuration relative to (c-2) due to chiral phonons with CW and CCW modes, respectively.}
\label{magnon_num}
\end{figure*}

In the case of ferromagnets, the expectation value of the total magnon number from the geometric terms becomes
\begin{align}
\label{Ntot}
N_{\rm{tot}}^{\rm{geom}}(\tau)=\sum_nN_n^{\rm{geom}}(\tau)=\Delta S^z(\tau),
\end{align}
which gives the change of electron spins. Figure~\ref{magnon_num} shows the expectation value of the number of magnons from the geometric term as a function of the dimensionless time $\tau$, and Fig.~\ref{magnon_num}(a-1)-(a-4) and Fig.~\ref{magnon_num}(b-1)-(b-4) are for the CW and CCW phonons, respectively. Here, $N_1^{\rm{geom}}$, $N_2^{\rm{geom}}$, and $N_3^{\rm{geom}}$ are the contributions from the first, the second, and the third bands, respectively, and their sum gives the total change of the spin $\Delta S^z=N_1^{\rm{geom}}+N_2^{\rm{geom}}+N_3^{\rm{geom}}$ in ferromagnets. The values of the middle numbers on the vertical axis showed in each figure represent their time average during a period of $2\pi$. We set the temperature $k_BT=2J_z$ for the Bose distribution function.

The results show that time averages of the number of magnons induced by the CW and CCW phonons have the same values but opposite signs for each band. The time-dependent Bloch Hamiltonians between the CW and CCW phonons can be connected by changing the sign of the dimensionless time $\tau$: $\mathcal{H}^{\rm{CW}}(\bm{k},\tau)=\mathcal{H}^{\rm{CCW}}(\bm{k},-\tau)$, because the CW and CCW phonons have the opposite directions of rotational motions and are mutually related via the time-reversal $\tau\rightarrow -\tau$. Hence the matrix elements of the number operator of magnons satisfy $\hat{N}^{\rm{CW}}_{nm}(\bm{k},\tau)=\hat{N}^{\rm{CCW}}_{nm}(\bm{k},-\tau)$, and the matrix elements of the Berry connection: $A^{\rm{CW}}_{mn}(\bm{k},\tau)=-A^{\rm{CCW}}_{mn}(\bm{k},-\tau)$. Therefore, the geometric terms of the number of magnons of the CW and CCW phonons are related by $N^{\rm{CW}}_n(\tau)=-N^{\rm{CCW}}_n(-\tau)$ in agreement with the numerical results in Fig.~\ref{magnon_num}(a-1)-(a-4) and Fig.~\ref{magnon_num}(b-1)-(b-4).

In the absence of chiral phonons, spin magnetization are fully polarized at absolute zero temperature $T=0$, in which magnons are not excited as shown in Fig.~\ref{magnon_num}(c-1). Once the temperature becomes finite but below the transition temperature $T_c$, magnons are excited, and spin magnetizations decrease with a precessional motion shown in Fig.~\ref{magnon_num}(c-2). Within our toy model, in the presence of the slow chiral phonons, the geometric effect gives a change of the number of magnons, and it corresponds to the time-dependent geometric term with a non-zero time average, which either increases or decreases the number of magnons. Figures~\ref{magnon_num}(d-1) and~\ref{magnon_num}(d-2) give schematic pictures of this process corresponding to our numerical calculations, in which the number of magnons increases with CW phonons and decreases with CCW phonons, respectively. It results in an opposite change of spin magnetization with an opposite atomic rotation reflecting the chiral nature of  phonons under an adiabatic process.

Changes of the number of magnons represents spin dynamics of electrons. $N_{\rm{tot}}^{\rm{geom}}(\tau)$ in Eq.~(\ref{Ntot}) means the expectation value of the magnon number per unit area in the $xy$ plane. Therefore, the magnon number per one localized spin is $N^{\rm{geom}}_{\rm{tot}}S_{\rm{uc}}/3$ with the size of the unit cell $S_{\rm{uc}}=\sqrt{3}a^2/2$ because the unit cell contains three spins. The numerical results in Fig.~\ref{magnon_num}(a-4) and~\ref{magnon_num}(b-4) show that $N_{\rm{tot}}^{\rm{geom}}$ is about $10^{-4}\times\hbar\omega a^2$, which means that the total magnon number per localized spin is about $10^{-5}$ if we take $\hbar\omega=0.1J_z=0.1$ as an example within the cope of adiabatic approximation. This value is to be compared with the total spin $S=1/2$ (see Eq.~(\ref{HP})). Thus, if we assume that the rotational radius of atoms around their equilibrium positions is about $10\%$ of the lattice constant within our model calculation, the change of the spin polarization due to the geometrical effect by the chiral phonons is about $10^{-5}$ of the full polarization in ferromagnets.

We comment that in reality the phonon amplitude is less than $10\%$, and the estimated magnitude of the effect will be smaller. On the other hand, within our theory because the effect is proportional to the phonon frequency (see Eq.~(\ref{geomN})), the effect will be much enhanced when the phonon frequency becomes larger. In our paper we take an adiabatic approximation, which is valid when $\hbar\omega/J_z\ll 1$, and we take $\hbar\omega/J_z=0.1$ in our calculation. Behaviors of this effect for larger values of the phonon frequency $\omega$ beyond the adiabatic approximation will be interesting and is left as a future work.

\section{conversion of chiral phonons into magnons in antiferromagnets} 
\label{secIII}
As ferromagnets hold spontaneous magnetization, the geometric effect of magnon excitations induced by chiral phonons in ferromagnets is not easy to distinguish from the equilibrium magnetization. In this section, we study the magnon excitations in antiferromagnets because antiferromagnets have no net magnetization in equilibrium. We will show that in the presence of chiral phonons, which modulate the spin-spin interaction in antiferromagnets, the net magnetization arises. This means that chiral phonons work as an effective magnetic field. In particular, here we introduce an antiferromagnet on the honeycomb lattice rather than the kagome lattice in order to avoid a frustrated spin configuration.
Meanwhile, a circular polarized phonon mode can also appear on a 2D honeycomb lattice, which has been already observed in experiments~\cite{CHexp}.

\subsection{Antiferromagnetic spin model on 2D honeycomb lattice}
We consider an antiferromagnetic spin model on the 2D honeycomb lattice consisting of two sublattices A and B as shown in Fig.~\ref{honeycomb}(a). The primitive vectors of the lattice are represented by $\bm b_1=b(1/2,\sqrt{3}/2)$ and $\bm b_2=b(-1/2,\sqrt{3}/2)$ with $b$ being the lattice constant, and the three nearest-neighbor vectors are labeled as $\bm d_1=b_0(0,-1)$, $\bm d_2=b_0(\sqrt{3}/2,1/2)$, and $\bm d_3=b_0(-\sqrt{3}/2,1/2)$ with $b_0=b/\sqrt{3}$ representing the length of the nearest-neighbor bond. We introduce an antiferromagnetic spin Hamiltonian given by
\begin{align}
\label{antiH0}
H_0=H_{\rm{ex}}+H_{\rm{DM}}+H_{\rm{mag}},
\end{align} 
where the first term
\begin{align}
H_{\rm{ex}}=\sum_{\braket{i,j}}\left(J_xS^x_iS^x_j+J_yS^y_iS^y_j+J_zS^z_iS^z_j\right)
\end{align}
is the anisotropic spin model with the exchange parameters $J_x>0$, $J_y>0$, and $J_z>0$ for antiferromagnets. The second term
\begin{align}
\label{honeyDM}
H_{\rm{DM}}=\sum_{\braket{\braket{i,j}}}\sum_{\mu=A,B}\bm D_{\mu}^{ij}\cdot(\bm S_{\mu,i}\times\bm S_{\mu,j})
\end{align}
is an out-of-plane next-nearest-neighbor DM interaction with $\bm D_{\mu}^{ij}=\nu_{\mu}^{ij}D_{\mu}\hat{\bm{e}_z}$, where $\nu_{\mu}^{ij}=\pm 1$ represents the direction from the $i$-th site to the $j$-th (the $j$-th site to the $i$-th) site along the dashed arrows shown in Fig.~\ref{honeycomb}(a) within the hexagonal plaquette~\cite{Moriya}, and the DM interactions for A and B sublattice are different in order to break the inversion symmetry. 
The third term is the same as Eq.~(\ref{Hmag}), which represents the coupling with an external magnetic field.

Here, we also take the $z$ axis as a quantization axis of electron spins similar to Sec.\ref{secII}. The antiferromagnetic spin Hamiltonian Eq.~(\ref{antiH0}) can be expressed in terms of ladder operators as
\begin{align}
\label{antispinH0}
H_0=&\frac{1}{4}\sum_{i,j}\Bigg\{4J_zS_i^zS_j^z+(J_x-J_y)(S_i^+S_j^++S_i^-S_j^-) \nonumber \\
&+(J_x+J_y)(S_i^+S_j^-+S_i^-S_j^+)-g\mu_B\tilde{H}\sum_iS_i^z \nonumber \\
&+\frac{1}{2}\sum_{\braket{\braket{i,j}}}\sum_{\mu=A,B}i\nu_{\mu}^{ij}(S_{\mu,i}^+S_{\mu,j}^--S_{\mu,i}^-S_{\mu,j}^+).
\end{align}
Since the spin configuration in antiferromagnets is anti-parallel between the two sublattices, the Holstein-Primakoff transformation for antiferromagnets can be defined as follows: for sublattice A,
\begin{align}
\label{HPA}
&S^z_{A,i}=S-b_{A,i}^{\dagger}b_{A,i}, \nonumber \\
&S^+_{A,i}=\left(2S-b_{A,i}^{\dagger}b_{A,i}\right)^{1/2}b_{A,i}\approx\sqrt{2S}b_{A,i}, \nonumber \\
&S^-_{A,i}=b_{A,i}^{\dagger}\left(2S-b_{A,i}^{\dagger}b_{A,i}\right)^{1/2}\approx\sqrt{2S}b_{A,i}^{\dagger}, 
\end{align}
and for sublattice B,
\begin{align}
\label{HPB}
&S^z_{B,i}=b_{B,i}^{\dagger}b_{B,i}-S, \nonumber \\
&S^+_{B,i}=b_{B,i}^{\dagger}\left(2S-b_{B,i}^{\dagger}b_{B,i}\right)^{1/2}\approx\sqrt{2S}b_{B,i}^{\dagger}, \nonumber \\
&S^-_{B,i}=\left(2S-b_{B,i}^{\dagger}b_{B,i}\right)^{1/2}b_{B,i}\approx\sqrt{2S}b_{B,i}, 
\end{align}
where $b_{A,i}^{\dagger}(b_{A,i})$ and $b_{B,i}^{\dagger}(b_{B,i})$ represents the creation (annihilation) operators of magnons for the sublattices A and B, respectively. By using this transformation, Eq.~(\ref{antispinH0}) is finally expressed as
\begin{align}
H_0=&\left(3J_zS-g\mu_B\tilde{H}\right)\sum_{i}\left(b_{A,i}^{\dagger}b_{A,i}+b_{B,i}^{\dagger}b_{B,i}\right) \nonumber \\
&+\frac{S}{2}(J_x-J_y)\sum_{\braket{i,j}}\left(b_{A,i}b_{B,i}^{\dagger}+b_{A,i}^{\dagger}b_{B,i}\right) \nonumber \\
&+\frac{S}{2}(J_x+J_y)\sum_{\braket{i,j}}\left(b_{A,i}b_{B,i}+b_{A,i}^{\dagger}b_{B,i}^{\dagger}\right) \nonumber \\
&+D_AS\sum_{\braket{\braket{i,j}}}i\nu_A^{ij}\left(b_{A,i}b_{A,j}^{\dagger}-b_{A,i}^{\dagger}b_{A,j}\right) \nonumber \\
&-D_BS\sum_{\braket{\braket{i,j}}}i\nu_B^{ij}\left(b_{B,i}b_{B,j}^{\dagger}-b_{B,i}^{\dagger}b_{B,j}\right),
\end{align}
where the number 3 represents the number of the nearest-neighbor atoms in 2D honeycomb lattice, and it can be obtained as a BdG formation as Eq.~(\ref{BdG}) by the Fourier transformation Eq.~(\ref{Fourier}). Here the basis $\beta^{\dagger}_{\bm k}=\left(b^{\dagger}_{A,\bm k},b^{\dagger}_{B,\bm k}\right)$ and the Bloch Hamiltonian
\begin{align}
\mathcal{H}_0(\bm k)=
\begin{pmatrix}
\mathcal{C}_{A,\bm k} & \Gamma_{\bm k} & 0 & \mathcal{D}_{\bm k} \\
\Gamma_{\bm k}^* & \mathcal{C}_{B,\bm k} & \mathcal{D}_{\bm k}^* & 0 \\
0 & \mathcal{D}_{-\bm k}^* & \mathcal{C}_{A,-\bm k} & \Gamma_{-\bm k}^* \\
\mathcal{D}_{-\bm k} & 0 & \Gamma_{-\bm k} & \mathcal{C}_{B,-\bm k}
\end{pmatrix},
\end{align} 
where $\mathcal{C}_{A,\bm k}=3J_zS+4D_ASg_{\bm k}$, 
$\mathcal{C}_{B,\bm k}=3J_zS+4D_BSg_{\bm k}$, 
$\Gamma_{\bm k}=\frac{S}{2}(J_x-J_y)f_{\bm k}$, and
$\mathcal{D}_{\bm k}=\frac{S}{2}(J_x+J_y)f_{\bm k}$
with $f_{\bm k}=\sin{(\bm k\cdot\bm b_1)}-\sin{(\bm k\cdot\bm b_2)}+\sin{[\bm k\cdot(\bm b_2-\bm b_1)]}$, and $g_{\bm k}=\sum_{j=1}^3e^{i\bm k\cdot\bm d_j}$. The magnon bands can be calculated by diagonalization of the BdG Hamiltonian placed in Appendix~\ref{A}, and its result has been shown in Fig.~\ref{honeycomb}(b) with parameters $J_z=1$, $J_x=0.7$, $J_y=0.4$, $D_A=0.2$, $D_B=0.1$, $S=1/2$ and $\tilde{H}=0$. There are two positive-definite bands above $E=0$ and two bands below $E=0$ are their copies.

\subsection{Modulation of spin-spin interactions by chiral phonons on 2D honeycomb lattice}
Next, we introduce a modulation of exchange parameters due to the chiral phonon at the $\bm \Gamma$ point of the phonon Brillouin zone for our discussions. At the $\bm \Gamma$ point, honeycomb lattice can have circular polarized phonon modes~\cite{PAM,CPthe,CHexp,CPmagHamada,CPcurrent}.
Here, we focus on the chiral phonon at the $\bm \Gamma$ point because chiral phonon at general non-$\bm \Gamma$ high-symmetry points can always be brought to the $\bm \Gamma$ point by appropriately enlarging the unit cell for honeycomb lattices.
 
\begin{figure}[htb]
\begin{center}
\includegraphics[clip,width=8.5cm]{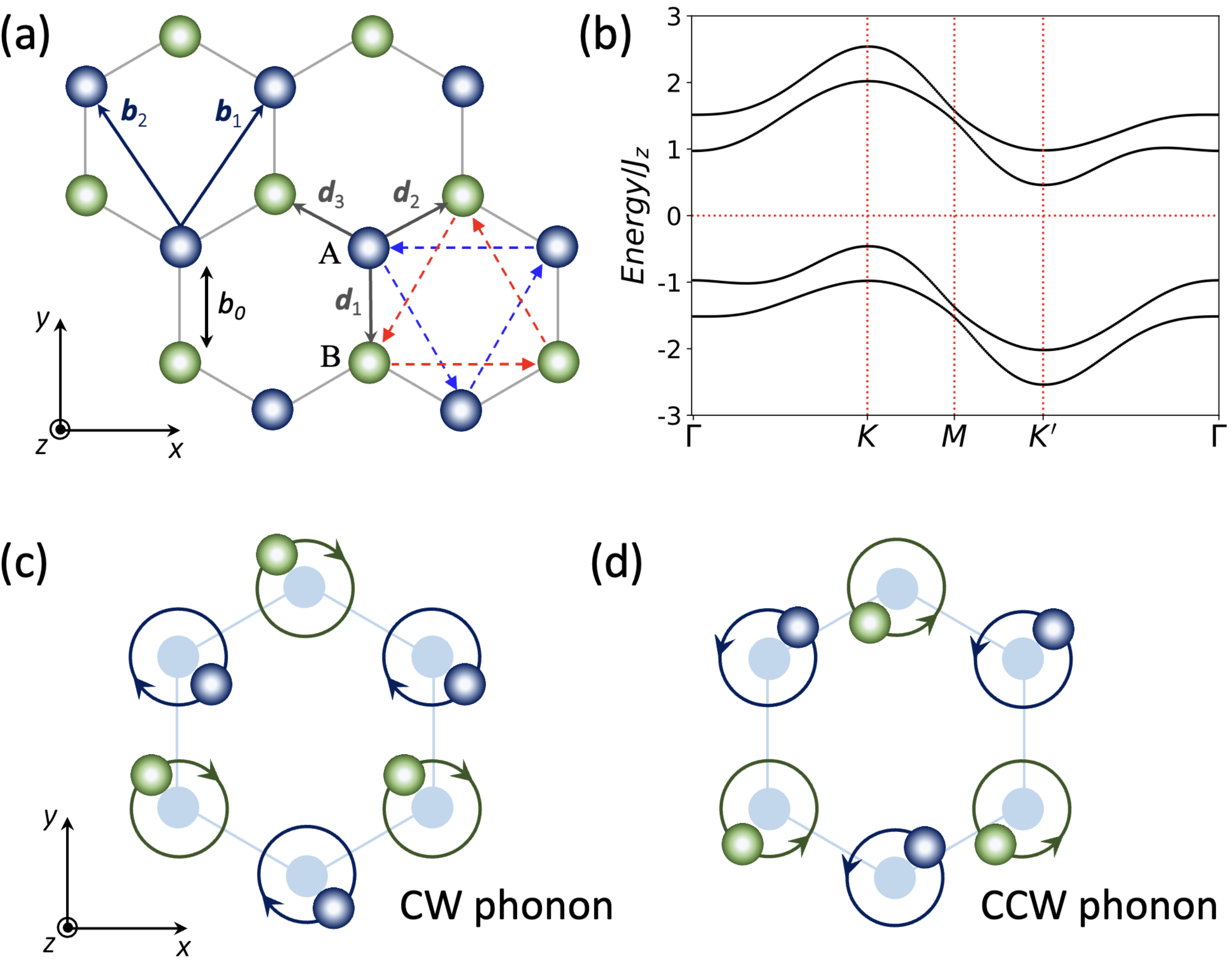}
\end{center}
\caption{
(Color online). (a) Two-dimensional honeycomb lattice with two sublattices A and B, and lattice vectors $\bm{b}_1$ and $\bm{b}_2$. Three nearest-neighbor vectors are labeled as $\bm d_i~(i=1,2,3)$. 
The dashed arrows give the positive $\nu_{ij}$ sign convention of DM interaction in Eq.~(\ref{honeyDM}).
(b) Magnon band structure with $J_z=1$, $J_x=0.7$, $J_y=0.4$, $D_A=0.2$, $D_B=0.1$, $S=1/2$ and $\tilde{H}=0$. There are two positive-defined bands above $E=0$, and two bands below $E=0$ are their copies.}
\label{honeycomb}
\end{figure}

In the case of the chiral phonon at the $\bm \Gamma$ point, we consider the optical branches, in which the same sublattice rotate with the same phase within the hexagonal plane, and the phase difference of the atomic rotation between the sublattices A and B is $\pi$. On the other hand, the atomic rotation around their equilibrium can be CW or CCW, which can be converted to each other via the time-reversal operation. Figures~\ref{honeycomb}(c) and~\ref{honeycomb}(d) show the schematics of the CW and CCW phonon modes. Here, the displacement vectors of the sublattice A and B for CW phonons shown in Fig.~\ref{honeycomb}(c) are written as $\bm r^{\rm{cw}}_A=r_0(\cos{\omega t},-\sin{\omega t})$ and $\bm r^{\rm{cw}}_B=-r_0(\cos{\omega t},-\sin{\omega t})$, and the sublattice A and B for CCW phonons shown in Fig.~\ref{honeycomb}(d) are written as $\bm r^{\rm{ccw}}_A=r_0(\cos{\omega t},\sin{\omega t})$ and $\bm r^{\rm{ccw}}_B=-r_0(\cos{\omega t},\sin{\omega t})$ with the phonon frequency $\omega$.
Since the length of the nearest-neighbor bond changes with time due to the chiral phonon, the exchange parameters are modulated as $J_{ij}^{\alpha}\rightarrow J_{ij}^{\alpha}-\frac{J_{ij}^{\alpha}}{b_0^2}\bm r(t)\cdot\bm d_{ij}~(\alpha=x,y)$ with the relative displacement $\bm r(t)=\bm r_B-\bm r_A=-2r_0(\cos{\omega t},-\sin{(\pm\omega t)})$ between the two sublattices, where $\pm$ represents the CW and CCW phonons, respectively. 

By using the Fourier transformation Eq.~(\ref{Fourier}), the modulated spin Hamiltonian can be written in a quadratic form of a bosonic operators as Eq.~(\ref{dHBdG}) with the basis $\beta^{\dagger}_{\bm k}=\left(b^{\dagger}_{A,\bm k},b^{\dagger}_{B,\bm k}\right)$:
\begin{align}
\delta\mathcal{H}(\bm k,t)=
\begin{pmatrix}
0 & \delta\Gamma_{\bm k}(t) & 0 & \delta\mathcal{D}_{\bm k}(t) \\
\delta\Gamma_{\bm k}^*(t) & 0 & \delta\mathcal{D}_{\bm k}(t) & 0 \\
0 & \delta\mathcal{D}_{-\bm k}^*(t) & 0 & \delta\Gamma_{-\bm k}^*(t) \\
\delta\mathcal{D}_{-\bm k}(t) & 0 & \delta\Gamma_{-\bm k}(t) & 0
\end{pmatrix}.
\end{align}
Here the elements of the modulated Bloch Hamiltonian are
$\delta\Gamma_{\bm k}(t)=S(\delta J_x-\delta J_y)h_{\bm k}(t)$ and 
$\delta\mathcal{D}_{\bm k}(t)=S(\delta J_x+\delta J_y)h_{\bm k}(t)$ with
$\delta J_{\alpha}=J_{\alpha}r_0/b_0~(\alpha=x,y)$ and
\begin{align}
h_{\bm k}(t)=&\sin{(\pm\omega t)}e^{i\bm k\cdot\bm d_1}+\left(\frac{\sqrt{3}}{2}\cos{\omega t}-\frac{1}{2}\sin{(\pm\omega t)}\right)e^{i\bm k\cdot\bm d_2} \nonumber \\
&+\left(-\frac{\sqrt{3}}{2}\cos{\omega t}-\frac{1}{2}\sin{(\pm\omega t)}\right)e^{i\bm k\cdot\bm d_3}, 
\end{align}
for CW and CCW phonons, respectively. Here $r_0$ denotes the amplitude of the atomic rotations on 2D honeycomb lattice, and we also assume that the amplitude is around 10$\%$ of the lattice constant $b_0$ of the honeycomb lattice with $r_0=0.1b_0$ in the following calculation as an example~\cite{CPmagHamada,CPcurrent}.

\subsection{Change of magnon numbers due to chiral phonons in antiferromagnets}
\begin{figure*}[htb]
\begin{center}
\includegraphics[clip,width=18.5cm]{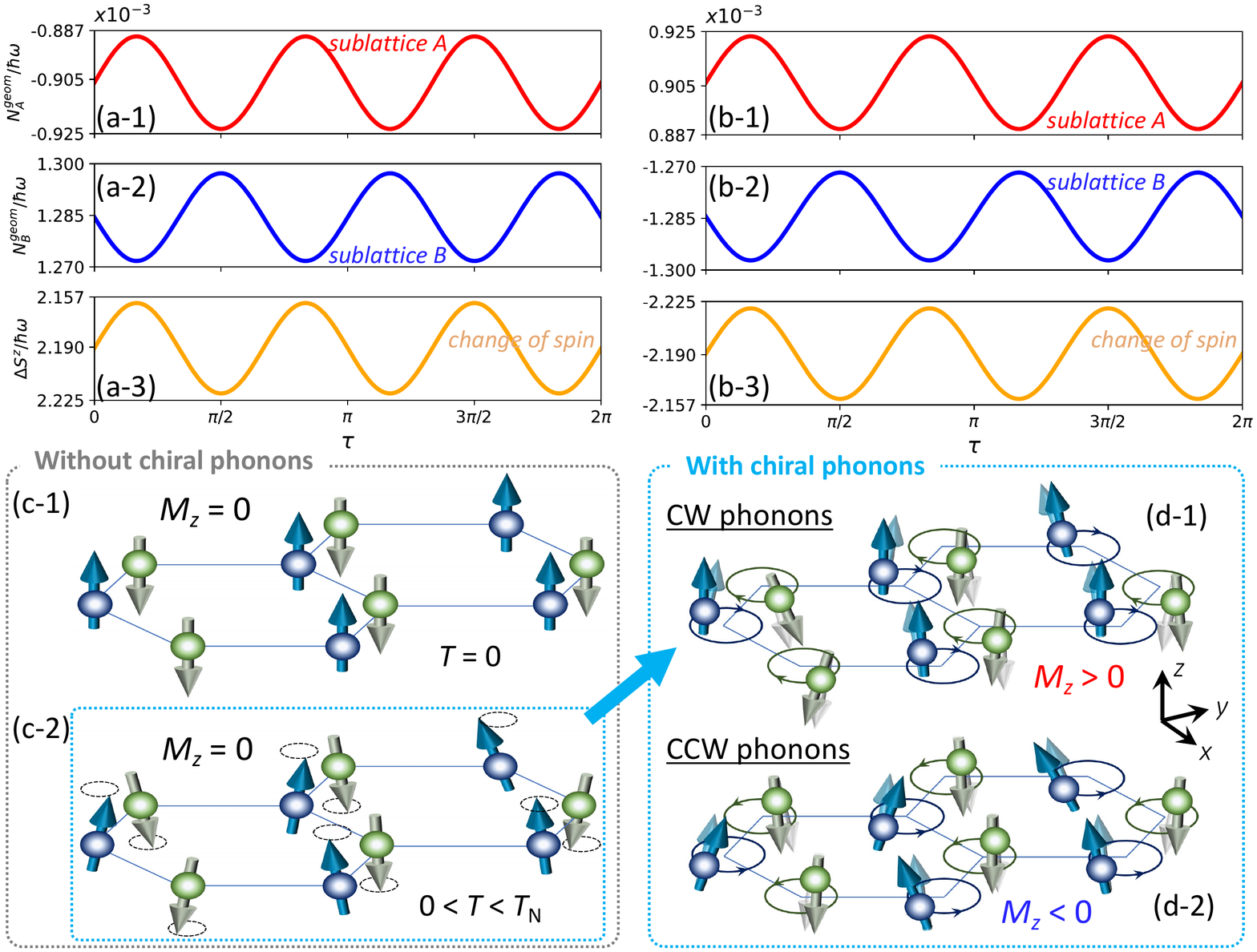}
\end{center}
\caption{
(Color online). The number of magnons varying with time $t$ for one period on 2D honeycomb lattice. Here $\tau=\omega t$ is the dimensionless time. (a)(b) Magnon excitation by (a) CW phonons and (b) CCW phonons. Here red and blue lines represent the number of magnons for the sublattice A and B, and orange lines represent the total change of spin $\Delta S=N_B-N_A$. The middle number on the vertical axis represents the time average over one period $0\leq\tau\leq2\pi$. The parameters are set to be $J_z=1$, $J_x=0.7$, $J_y=0.4$, $D_A=0.2$, $D_B=0.1$, $r_0=0.1b_0$, $S=1/2$, and $\tilde{H}=0$. (c-1) and (c-2) show schematic figures of spin configuration without chiral phonons, in which the net magnetization is zero below the Neel temperature $T_N$. (d-1) and (d-2) show the changed spin configuration relative to (c-2) due to chiral phonons with CW and CCW modes, respectively. The net magnetization becomes nonzero due to chiral phonons.}
\label{anti_magnon_num}
\end{figure*}

As we discussed in Sec.~\ref{secII}, geometric effect during the adiabatic process gives an additional magnon excitation due to chiral phonons, in which the number of magnons has a non-trivial contribution dubbed geometric term. In the case of the antiferromagnetic honeycomb lattice, the creation (annihilation) of magnons on the sublattices A and B gives the opposite changes of spin magnetization $S^z$. Here we set that the spin $S^z_A$ on the sublattice A decreases (increases) and the spin $S^z_B$ on the sublattice B increases (decreases). In the presence of the chiral phonon, the geometric term of the magnon number for each sublattice changes with time. Figure~\ref{anti_magnon_num}(a-1)(a-2) and~\ref{anti_magnon_num}(b-1)(b-2) show our numerical results for the magnon numbers with CW and CCW phonons, respectively. Here, $N^{\rm{geom}}_A(\tau)$ and $N^{\rm{geom}}_B(\tau)$ represent the magnon number of sublattice A and B. Moreover, the total change of spin magnetization $\Delta S^z(\tau)=N^{\rm{geom}}_A(\tau)-N^{\rm{geom}}_B(\tau)$~(see Eq.~(\ref{HPA}) and Eq.~(\ref{HPB})) is shown in Fig.~\ref{anti_magnon_num}(a-3) and~\ref{anti_magnon_num}(b-3) for CW and CCW phonons. The values of the middle number on the vertical axis showed in each figure represent the time average during a period of $2\pi$ for $\tau$. We set the temperature $k_BT=2J_z$ for the Bose distribution function.

The results show that time averages of the change in the number of magnons induced by the CW and CCW phonons have the same magnitudes but opposite signs. 
Since the CW and CCW phonons have the opposite directions of rotational motions and are mutually related via the time-reversal $\tau\rightarrow -\tau$, and their Hamiltonians statisfy $\mathcal{H}^{\rm{CW}}(\bm{k},\tau)=\mathcal{H}^{\rm{CCW}}(\bm{k},-\tau)$, which gives the matrix elements of the number operator of magnons: $\hat{N}^{\rm{CW}}_{nm}(\bm{k},\tau)=\hat{N}^{\rm{CCW}}_{nm}(\bm{k},-\tau)$, and the matrix elements of the Berry connection: $A^{\rm{CW}}_{mn}(\bm{k},\tau)=-A^{\rm{CCW}}_{mn}(\bm{k},-\tau)$. Therefore, the geometric terms of the number of magnons of the CW and CCW phonons are related by $N^{\rm{CW}}_n(\tau)=-N^{\rm{CCW}}_n(-\tau)$ in agreement with the numerical results in Fig.~\ref{anti_magnon_num}(a-1)-(a-3) and Fig.~\ref{anti_magnon_num}(b-1)-(b-3).

Without chiral phonons, antiferromagnets hold a spin configuration with neighboring spins on different sublattices pointing in opposite directions below the $\rm{N\acute{e}el}$ temperature $T_N$, which corresponds to a vanishing net magnetization shown in Fig.~\ref{anti_magnon_num}(c-1)(c-2). On the other hand, chiral phonons with atomic rotations work as an effective magnetic field, which results in a non-zero net magnetization in antiferromagnets. In the presence of slow chiral phonons, the geometric effect changes spin magnetization from zero to a finite value, with opposite signs between the CW and CCW phonons as schematically shown in Fig.~\ref{anti_magnon_num}(d-1)(d-2). The numerical results in Fig.~\ref{anti_magnon_num}(a-3) and~\ref{anti_magnon_num}(a-4) show that change of spin magnetization $\Delta S^z$ per unit area in the $xy$ plane from the geometric effect is about $10^{-3}\times\hbar\omega a^2$, which means that the net magnetization is about $10^{-4}$ compared with one local spin $S=\pm1/2$ at the absolute temperature $T=0$~(see Eq.~(\ref{HPA}) and Eq.~(\ref{HPB})) if we take $\hbar\omega=0.1J_z=0.1$ as an example within adiabatic approximation.

We comment that in reality the phonon amplitude is less than 10$\%$, and the estimated magnitude of the effect will be smaller. On the other hand, within our theory because the effect is proportional to the phonon frequency (see Eq.~(\ref{geomN})), the effect will be much enhanced when the phonon frequency becomes larger. As the net magnetization becomes finite from zero due to chiral phonons in antiferromagnets, the change of magnetization can be observed more easily by experiments than that in ferromagnets.

\section{conclusion}
\label{secIV}
In this paper, we have theoretically proposed a new conversion of chiral phonons into magnons in ferromagnets and antiferromagnets. We construct a ferromagnetic spin model on a 2D kagome lattice and an antiferromagnetic spin model on a 2D honeycomb lattice with anisotropic exchange and DM interactions to study the geometric magnon excitation generated by chiral phonons. We find that it is necessary to introduce anisotropy of the spin-spin interaction. Namely we need to break spin-rotation symmetry with respect to the $z$ direction in order to obtain a non-trivial change of spin magnetization.  Our model is intended to be a minimal model to demonstrate a new type of magnon excitations induced by slow chiral phonons from the geometric effects.

By treating the atomic rotations as an adiabatic process, we find that the time-dependent number of magnons has two terms: the instantaneous term and the geometric term. 
The instantaneous term is trivially given by a snapshot at the given time, so the instantaneous term will be of no difference between the CW and CCW phonon modes. 
On the other hand, for the geometric term, intriguingly the time averages of the number of magnons induced by the CW and CCW phonons have the same magnitudes but opposite signs. Thus, the phonon chirality is reflected by the geometric effect here.
As a result, due to the geometric effect of chiral phonons, the spin magnetization changes with time. We can interpret this phenomenon as a conversion of chiral phonons into magnons both in ferromagnets and antiferromagnets. While it may not be easy to measure this effect in ferromagnets due to the presence of uniform magnetization, measurement in antiferromagnets is easier because the net spin magnetization becomes nonzero due to chiral phonons working as an effective magnetic field. 

Different from the previous studies related to phonon-magnon coupling~\cite{ph-ma-coup,ChMo}, in our model the contribution from chiral phonons is absorbed as time-dependent parameters into the spin Hamiltonian. Here, even though the conversion of chiral phonons into magnons is taken into account, we can still simply deal with it as a single-particle problem for magnons. Thus, the similar method can also be used to investigate general systems with slow chiral phonons. 
On the other hand, for phonons with higher frequency, the effect predicted in this paper will become larger but its calculation is beyond the scope of the Berry phase method we used in the present paper. Such behaviors of the chiral phonon with higher frequency are left as future problems.

\begin{acknowledgments}

This work was partly supported by JSPS KAKENHI Grants No. JP20H04633, No. JP22H00108, and No. JP23KJ0926, and also by MEXT Initiative to Establish Next-generation Novel Integrated Circuits Centers (X-NICS) Grant No. JPJ011438.
\end{acknowledgments}

\begin{appendix}

\section{Diagonalization of bosonic BdG Hamiltonian}
\label{A}
In this appendix, we provide more details about the diagonalization of a generic bosonic BdG Hamiltonian~\cite{BdG1,BdG2}.
As we showed in the main text, the generic bosonic BdG Hamiltonian can be expressed as
\begin{align}
\label{quadratic}
H=\frac{1}{2}\sum_{\bm{k}}
\begin{pmatrix}
\bm{\beta}_{\bm{k}}^{\dagger}, & \bm{\beta}_{-\bm{k}}
\end{pmatrix}
\mathcal{H}({\bm{k}})
\begin{pmatrix}
\bm{\beta}_{\bm{k}} \\
\bm{\beta}_{-\bm{k}}^{\dagger}
\end{pmatrix},
\end{align}
with $\bm{\beta}_{\bm{k}}^{\dagger}\equiv \left(a_{1,\bm{k}}^{\dagger},a_{2,\bm{k}}^{\dagger},\cdots,a_{M,\bm{k}}^{\dagger}\right)$, $a_{j,\bm{k}}^{\dagger} (j=1,2,...,M)$ is a boson creation operator, and $M$ represents the internal degrees of freedom within a unit cell. The Bloch Hamiltonian $\mathcal{H}(\bm{k})$ is a $2M\times2M$ Hermitian matrix, and is generally expressed as
\begin{align}
\mathcal{H}(\bm{k})=
\begin{pmatrix}
\mathcal{A}_{\bm{k}} & \mathcal{B}_{\bm{k}} \\
\mathcal{B}^*_{-\bm{k}} & \mathcal{A}^*_{-\bm{k}} \\
\end{pmatrix},
\end{align}
with $M\times M$ matrices $\mathcal{A}_{\bm{k}}$ and $\mathcal{A}^*_{-\bm{k}}$ for the normal channel, and $M\times M$ matrices $\mathcal{B}_{\bm{k}}$ and $\mathcal{B}^*_{-\bm{k}}$ for the anomalous channel.
To diagonalize such a bosonic BdG Hamiltonian, we should introduce a paraunitary matrix $\mathcal{T}(\bm{k})$ instead of a unitary matrix~\cite{BdGdiag},
\begin{align}
\label{parau}
\mathcal{T}^{\dagger}(\bm{k})\mathcal{H}(\bm{k})\mathcal{T}(\bm{k})=
\begin{pmatrix}
E_{\bm{k}} &  \\
 & E_{-\bm{k}} \\
\end{pmatrix},
\end{align}   
where $E_{\bm{k}}$ is a $M\times M$ diagonal matrix with its diagonal elements representing the magnon energy eigenvalues. The paraunitary matrix $\mathcal{T}(\bm{k})$ changes the old basis $\left(\bm{\beta}^{\dagger}_{\bm{k}},\bm{\beta}_{-\bm{k}}\right)$ into a new basis $\left(\bm{\gamma}^{\dagger}_{\bm{k}},\bm{\gamma}_{-\bm{k}}\right)$ as
\begin{align}
\label{basistran}
\left(\bm{\gamma}^{\dagger}_{\bm{k}},\bm{\gamma}_{-\bm{k}}\right)\mathcal{T}^{\dagger}(\bm{k})=\left(\bm{\beta}^{\dagger}_{\bm{k}},\bm{\beta}_{-\bm{k}}\right),
 \end{align}
and it satisfies 
\begin{align}
\label{Tsigma}
\mathcal{T}^{\dagger}(\bm{k})\Sigma_3\mathcal{T}(\bm{k})=\Sigma_3,~\mathcal{T}(\bm{k})\Sigma_3\mathcal{T}^{\dagger}(\bm{k})=\Sigma_3,
\end{align}
with a diagonal $2M\times 2M$ diagonal matrix $\Sigma_3=s_z\otimes I_{M\times M}$, where $s_z$ is the $z$ component of the $2\times 2$ Pauli matrices, and $I_{M\times M}$ is a $M\times M$ identity matrix. 
The Hermitian BdG Hamiltonian $\mathcal{H}(\bm{k})$ should be a positive-definite matrix in order to ensure that the magnon excitation energies are positive for any $\bm{k}$.

The paraunitary matrix $\mathcal{T}(\bm{k})$ used to diagonalize $\mathcal{H}(\bm{k})$ can be obtained by the following procedure. First, we decompose $\mathcal{H}(\bm{k})$ into a product between an upper triangle matrix $\mathcal{K}(\bm{k})$ and its Hermitian conjugate, $\mathcal{H}(\bm{k})=\mathcal{K}^{\dagger}(\bm{k})\mathcal{K}(\bm{k})$ by the Cholesky decomposition. The Cholesky decomposition is possible as long as the BdG Hamiltonian $\mathcal{H}(\bm{k})$ is positive definite.
Then we diagonalize a matrix $\mathcal{W}(\bm{k})\equiv\mathcal{K}(\bm{k})\Sigma_3\mathcal{K}^{\dagger}(\bm{k})$ as 
\begin{align}
\mathcal{U}(\bm{k})^{\dagger}\mathcal{W}(\bm{k})\mathcal{U}(\bm{k})=
\begin{pmatrix}
E_{\bm{k}} & \\
 & -E_{-\bm{k}} \\
\end{pmatrix},
\end{align} 
by introducing a unitary matrix $\mathcal{U}(\bm{k})$. Here, $E_{\bm{k}}$ is an $M\times M$ diagonal matrix with positive diagonal elements. Finally, the paraunitary matrix is obtained as
\begin{align}
\mathcal{T}(\bm{k})=\mathcal{K}^{-1}(\bm{k})\mathcal{U}(\bm{k})
\begin{pmatrix}
E^{1/2}_{\bm{k}} &  \\
 & E^{1/2}_{-\bm{k}} \\
\end{pmatrix},
\end{align}
which satisfies Eq.~(\ref{Tsigma}). Thus, by means of the Cholesky decomposition, we can obtain the magnon bands from the bosonic BdG Hamiltonian.

\section{Geometric term of in bosonic BdG formalization}
\label{B}
In this appendix, we explain how to calculate the geometric term of the number of magnons $\hat{N}$ for the bosonic BdG formalization mentioned in the main text.
In Appendix~\ref{A}, we showed that a generic BdG bosonic Hamiltonian Eq.~(\ref{quadratic}) is written as the quadratic form in the basis $\left(\bm{\beta}^{\dagger}_{\bm{k}}, \bm{\beta}_{-\bm{k}}\right)$, and meanwhile the matrix of the number of magnons $\hat{N}$ in the same basis is an identity matrix which is expressed as
\begin{align}
\hat{N}=\frac{1}{2}\sum_{\bm{k}}
\begin{pmatrix}
\bm{\beta}_{\bm{k}}^{\dagger}, & \bm{\beta}_{-\bm{k}}
\end{pmatrix}
\begin{pmatrix}
\bm{\beta}_{\bm{k}} \\
\bm{\beta}_{-\bm{k}}^{\dagger}
\end{pmatrix}.
\end{align}
The BdG Bloch Hamiltonian can be diagonalized by the basis transformation of Eq.~(\ref{basistran}) to $\left(\bm{\gamma}^{\dagger}_{\bm{k}},\bm{\gamma}_{-\bm{k}}\right)$, and the time-dependent matrix of the BdG Bloch Hamiltonian can be written as
\begin{align}
\hat{H}(\tau)=&\frac{1}{2}\sum_{\bm{k}}
\begin{pmatrix}
\bm{\beta}_{\bm{k}}^{\dagger}(\tau), & \bm{\beta}_{-\bm{k}}(\tau)
\end{pmatrix}
\mathcal{H}({\bm{k}},\tau)
\begin{pmatrix}
\bm{\beta}_{\bm{k}}(\tau) \\
\bm{\beta}_{-\bm{k}}^{\dagger}(\tau)
\end{pmatrix} \nonumber \\
=&\frac{1}{2}\sum_{\bm{k}}
\begin{pmatrix}
\bm{\gamma}_{\bm{k}}^{\dagger}, & \bm{\gamma}_{-\bm{k}}
\end{pmatrix}
\begin{pmatrix}
E_{\bm{k}}(\tau) &  \\
 & E_{-\bm{k}}(\tau)
\end{pmatrix}
\begin{pmatrix}
\bm{\gamma}_{\bm{k}} \\
\bm{\gamma}_{-\bm{k}}^{\dagger}
\end{pmatrix},
\end{align}
which becomes diagonal and the time dependence is reduced into the diagonal matrix. Here $E_{\bm{k}}$ is a $M\times M$ matrix whose diagonal elements show the particle dispersion of bosons. Meanwhile, the number of magnon becomes
\begin{align}
\hat{N}(\tau)=\frac{1}{2}\sum_{\bm{k}}
\begin{pmatrix}
\bm{\gamma}_{\bm{k}}^{\dagger}, & \bm{\gamma}_{-\bm{k}}
\end{pmatrix}
\mathcal{T}^{\dagger}(\bm{k},\tau)\mathcal{T}(\bm{k},\tau)
\begin{pmatrix}
\bm{\gamma}_{\bm{k}} \\
\bm{\gamma}_{-\bm{k}}^{\dagger}
\end{pmatrix}.
\end{align}
Therefore, in Eq.~(\ref{geomN}), the matrix elements of the number operator of magnons (Eq.~(\ref{Nnm})) and those of the Berry connection (Eq.~(\ref{Amn})) are rewritten as
\begin{align}
\hat{N}_{nm}(\bm{k},\tau)=\frac{1}{2}\bigg\{\mathcal{T}^{\dagger}(\bm{k},\tau)\mathcal{T}(\bm{k},\tau)\bigg\}_{nm},
\end{align}
and
\begin{align}
A_{mn}(\bm{k},\tau)=-i\frac{\bigg\{\mathcal{T}^{\dagger}(\bm{k},\tau)\partial_{\tau}\mathcal{H}(\bm{k},\tau)\mathcal{T}(\bm{k},\tau)\bigg\}_{mn}}{E_{n,\bm{k}}(\tau)-E_{m,\bm{k}}(\tau)},
\end{align}
respectively.

\end{appendix}

\end{document}